\documentclass[preprint,10pt]{article}
\usepackage{amssymb}
\usepackage{caption}
\usepackage{breqn}
\usepackage{graphicx}

\usepackage{caption}
\usepackage{subcaption}
\usepackage{physics}
\usepackage{multirow}
 \usepackage[numbers, sort&compress]{natbib} 

\begin{document}

\begin{center}
\textbf{KNO scaling, memorylessness and  maximal entanglement at  universal fixed point}

 Mustapha Ouchen and Alex Prygarin \\
\small
   Department of Physics, Ariel University, Ariel 40700, Israel
\end{center}
\normalsize

\normalsize

\begin{abstract}

 We analyze the experimental data of $\mathtt{p}\mathtt{-}\mathtt{p}$ collisions by the ATLAS and confront it with the AGK model developed by two of the authors,   the Kharzeev-Levin~(KL) model and the simple exponential behavior for the Koba, Nielsen and Olesen~(KNO) scaling function. We show that all three models virtually coincide with all available  experimental curves crossed  at value $z=n/\langle n \rangle$ of the KNO scaling parameter in the  broad range of the center-of-mass energies.
The theoretical results and the experimental data show  that the KNO violating terms are highly suppressed  in the vicinity of $z=2$. 
The special status of the universal scaling point $z=2$, where the KNO scaling is restored for $\mathtt{p}\mathtt{-}\mathtt{p}$ collisions, suggests that it originates from statistical saturation of the optical theorem. 
  We claim that any unitary model that approximates memoryless distribution in the high energy limit  must  have the same KNO scaling function $e^{-z}$ in the vicinity of  $z=2$ with KNO violating corrections of the order of one over average multiplicity squared.  This reflects the maximal entanglement of the final states.

\end{abstract}

\section{Introduction}

In this paper we  address the recent developments in applying the quantum entanglement to describing hadronization  in the high energy scattering~\cite{Baker:2017wtt,Kharzeev:2021nzh,Hentschinski:2022rsa,Kutak:2023cwg,Hentschinski:2023nhy,Hentschinski:2023izh,Hentschinski:2024gaa,Grieninger:2023ufa,Chachamis:2023omp,Bhattacharya:2024sno,Ouchen:2025tta,Grieninger:2025wxg,Datta:2024hpn,Hentschinski:2025pyq,Kutak:2025hzo,Kutak:2025syp,Fucilla:2025kit,Moriggi:2025qfs,Hatta:2024lbw,Ramos:2025tge,Moriggi:2024tiz}  and we discuss  the Koba, Nielsen and Olesen~(KNO)~\cite{Polyakov:1970lyy, Koba:1972ng} scaling hypothesis  for  $p$$-$$p$ collisions and show  that there is a special region, where the KNO violating corrections are strongly suppressed and the KNO scaling is valid with accuracy well beyond experimental errors.  
The Koba, Nielsen and Olesen~(KNO) scaling hypothesis states  that the multiplicity distribution must scale as 
\begin{eqnarray}
P_n(\sqrt{s})= \frac{1}{\langle n\rangle} \Psi \left(z \right)+\mathcal{O}\left( \frac{1}{\langle n\rangle^2} \right), 
\label{KNOdef}
\end{eqnarray}
where $\langle n\rangle$ is the mean multiplicity, the scaling parameter $z=n/\langle n\rangle$ and $\Psi(z)$ is some smooth analytic function. 
It is known~\cite{ UA5_1986, CMS_2011, ATLAS_2022, ALICE_2010, Sjostrand_1987,Kulchitsky:2022gkm,Kulchitsky:2023fqd} that the KNO scaling is violated for $\mathtt{p}-\mathtt{p}$ collision, where the expression $ \langle n\rangle P_n(\sqrt{s}) $ shows strong energy dependence instead of being a function of the  only    variable $z$. 
We calculate the KNO function $\Psi \left(z \right)$ using the AGK model developed by the authors~\cite{Ouchen:2025tta} using the Abramovski-Gribov-Kancheli~(AGK) cutting rules~\cite{AGK_1973, Baker_1976, Bartels_2006} applied to the Markov chain~\cite{Levin:2008cgz, Mueller:1985wy, Salvadore:2007th, Gelis:2006ye}
  and compare it to the experimental data for $p$$-$$p$ collisions.  
We show that the AGK model is very  good in describing the experimental data in a wide range of the center-of-mass energies and  transverse momenta. The high energy expansion of   the AGK model reveals a special point of $z=2$, where the higher order corrections are strongly suppressed compared to other values of $z$. We find the experimental evidence for this point and demonstrate that all experimental curves of the KNO scaled function  $\langle n\rangle \Psi \left(z \right)$ cross at $z=2$ 
for $\sqrt{s}=0.9, \; 2.36, \; 7, \; 8, \; 13 \; \mathtt{TeV}$. In the vicinity of $z=2$  the KNO scaled function in the AGK model and the Kharzeev-Levin model~\cite{Kharzeev:2017qzs} can be reasonably  approximated  by the exponential function $e^{-z}$. We argue that this fact follows from  the memorylessness  property of  the produced particles   distribution
described by the Markov chain in the AGK model and the KL model. We argue that the universal scaling point $z=2$, at which the number of produced 
particles equals twice the average multiplicity $n=2\langle n\rangle$,  is related to  the optical theorem and should hold for any distribution that respects the unitarity condition provided the initial condition remains unchanged. The simple exponential behavior of the KNO scaled function at the vicinity of the universal scaling point $z=2$ implies the maximal entanglement  of the final states in this region.  

%



\section{KNO function in  KL and AGK models}
In this section we briefly review results of the recent study of two of the authors~\cite{Ouchen:2025tta}, where the Abramovskii-Gribov-Kancheli~(AGK) cutting rules were used to derive 
the entanglement entropy for the evolution of cut and uncut pomerons. This continues a series of recent works~\cite{Kharzeev:2021nzh,Hentschinski:2022rsa,Kutak:2023cwg,Hentschinski:2023nhy,Hentschinski:2023izh,Hentschinski:2024gaa,Grieninger:2023ufa,Chachamis:2023omp}  initiated by Kharzeev and Levin~\cite{Levin:2008cgz}~(the KL model)  applying Markov chains to the study of the quantum
entanglement in the high energy scattering. The resulting  AGK model for the pomeron evolution in  the zero transverse dimension was successful  in describing the experimental data for $q$-moments   $C_q$ in the wide range of the center-of-mass energies without use of any adjustable parameter. 

The AGK model~\cite{Ouchen:2025tta} results in  a probability  given by
\begin{eqnarray}
P^{AGK}_n(Y)=\left(1- \frac{2^n}{3^n}\right)  \left(1-e^{-\alpha  Y }\right)^{n-1}e^{-\alpha Y} \left(1+2 e^{-\alpha  Y}\right), \;\;
\label{PnAGK}
\end{eqnarray}
where $Y$ is the rapidity and  $\alpha$ is the coupling constant.  $P_n(Y)$ in \eqref{PnAGK} describes the probability of having $n$  pomerons under condition that at least one of them is the cut pomeron. 
This should be compared to the KL model 
\begin{eqnarray}
P^{KL}_n(Y)=  \left(1-e^{-\alpha  Y }\right)^{n-1}e^{-\alpha Y} , \;\;
\label{PnKL}
\end{eqnarray}
where no distinction is made between cut and uncut pomerons. 
The Koba, Nielsen and Olesen~(KNO) scaling hypothesis states  that the multiplicity distribution must scale as 
\begin{eqnarray}
P_n(\sqrt{s})= \frac{1}{\langle n\rangle} \Psi \left(z \right)+\mathcal{O}\left( \frac{1}{\langle n\rangle^2} \right), 
\label{KNOdef2}
\end{eqnarray}
and it is convenient to consider 
\begin{eqnarray}
 \langle n\rangle  P_n(Y)
\end{eqnarray}
which equals the KNO function $\Psi \left(z \right)$ up to sub-leading corrections as  follows from (\ref{KNOdef2}).
The average $\langle n\rangle$ for  the KL model reads  
\begin{eqnarray}
\langle n\rangle^{KL}=e^{\alpha  Y }
\label{aveKL}
\end{eqnarray}
and    for the AGK models the average is given by 
\begin{eqnarray}
\langle n\rangle^{AGK}=e^{\alpha  Y}-1+\frac{3}{2 e^{-\alpha  Y}+1}.
\label{aveAGK}
\end{eqnarray}
We express  $e^{\alpha Y}$  in terms of $\langle n\rangle$  and then expand $ \langle n\rangle  P_n(Y)$ for large $\langle n\rangle$ to the second order. The resulting expressions for the KL model 
\begin{eqnarray}
 \left(\langle n\rangle  P_n(Y)\right)^{AGK}  &\simeq  & e^{-z} \left\{1-\frac{5 (z-2)}{2 \langle n\rangle}+\frac{75 z^2-308 z+312}{24 \langle n\rangle^2} 
  \right. \nonumber
  \\
  && -
\left. \left(\frac{2}{3}\right)^{z  \langle n\rangle  } \left(1-\frac{5 (z-2)}{2 \langle n\rangle}  \right) \right\}
\end{eqnarray}
and the AGK model 
\begin{eqnarray}
 \left(\langle n\rangle  P_n(Y)\right)^{KL}  \simeq e^{-z} \left( 1-\frac{z-2}{2  \langle n\rangle} +\frac{3 z^2-20 z+24}{24  \langle n\rangle^2}\right)
\end{eqnarray}
show a common feature of the vanishing first order KNO  violating correction    at $z=2$. 

It is instructive to perform this expansion at the vicinity of $z=2$ for experimentally relevant values of $\langle n\rangle$ ranging from $6.53$ to $33.88$ as given in Table~\ref{table:aveValues} for  the average multiplicity of the produced particles.  \ref{table:aveValues}
\begin{table}[!ht]
    \centering
     \begin{tabular}{|c|c|c|}
\hline
$\sqrt{s}$ & $p_{\text{T}}^{\min}$ & Average \\
$[\text{TeV}]$ & $[\text{MeV}]$ & Multiplicity \\
\hline
\multirow{2}{*}{13} & 100 & $33.88 \pm 0.11$ \\
\cline{2-3}
                    & 500 & $14.66 \pm 0.04$ \\
\hline
\multirow{2}{*}{8}  & 100 & $29.81 \pm 0.10$ \\
\cline{2-3}
                    & 500 & $12.25 \pm 0.03$ \\
\hline
\multirow{2}{*}{7}  & 100 & $29.40 \pm 0.19$ \\
\cline{2-3}
                    & 500 & $11.98 \pm 0.05$ \\
\hline
2.36                & 500 & $8.66 \pm 0.51$ \\
\hline
\multirow{2}{*}{0.9}& 100 & $18.06 \pm 0.12$ \\
\cline{2-3}
                    & 500 & $6.53 \pm 0.03$ \\
\hline 
\end{tabular}
\caption{The average multiplicity calculated by Kulchitsky and Tsiareshka~\cite{Kulchitsky:2022gkm} fitting the data of  ATLAS
Collaboration  for  the  pseudorapidity region  $|\eta| < 2.5 $ and 
 the center-of-mass energies  $\sqrt{s}= 0.9 \;\mathtt{TeV}, 7 \;\mathtt{TeV}, 8 \;\mathtt{TeV}, 13 \;\mathtt{TeV}$. }
 \label{table:aveValues}
\end{table}
The KL and the AGK model give similar results at $z=2$
\begin{eqnarray}
\left(\langle n\rangle  P_n(Y)\right)^{AGK}|_{z=2} \simeq e^{-2} \left(1-\frac{1}{6 \langle n\rangle^2} -\left(\frac{2}{3}\right)^{2 \langle n\rangle}\right)
\end{eqnarray}
and 
\begin{eqnarray}
\left(\langle n\rangle  P_n(Y)\right)^{KL}|_{z=2} \simeq e^{-2} \left(1-\frac{1}{6 \langle n\rangle^2} \right)
\end{eqnarray}

The difference term $\left(\frac{2}{3}\right)^{2 \langle n\rangle}$ can be safely neglected for $\langle n\rangle> 10$, i.e. for the experimental data $\sqrt{s}= 7 \;\mathtt{TeV}, 8 \;\mathtt{TeV}, 13 \;\mathtt{TeV}$. The KL model and the AGK  model   both successfully describe the experimental data for $p_{\text{T}}^{\min} >500 \;\mathtt{MeV}$ as will be shown later. 

The expansion of the KL model and the AGK model for large $\langle n \rangle$ shows that $z=2$ is a very special point where the first order corrections to the leading exponential behaviour $\Psi_{KNO} \simeq e^{-z}$ vanish. The universality of the $z=2$ point found in the two independent models suggests that such a point should be seen in the experimental data and this fact should be model independent.   Our analysis shows the same results for our models~\cite{Kharzeev:2021nzh,Hentschinski:2022rsa,Kutak:2023cwg,Hentschinski:2023nhy,Hentschinski:2023izh,Hentschinski:2024gaa,Grieninger:2023ufa,Chachamis:2023omp} obtained by either modifications of the KL model or QCD inspired calculations. 

\begin{figure*}[htb!]
\begin{minipage}[h]{0.50\textwidth} 
\center{\includegraphics[width=1.0\linewidth]{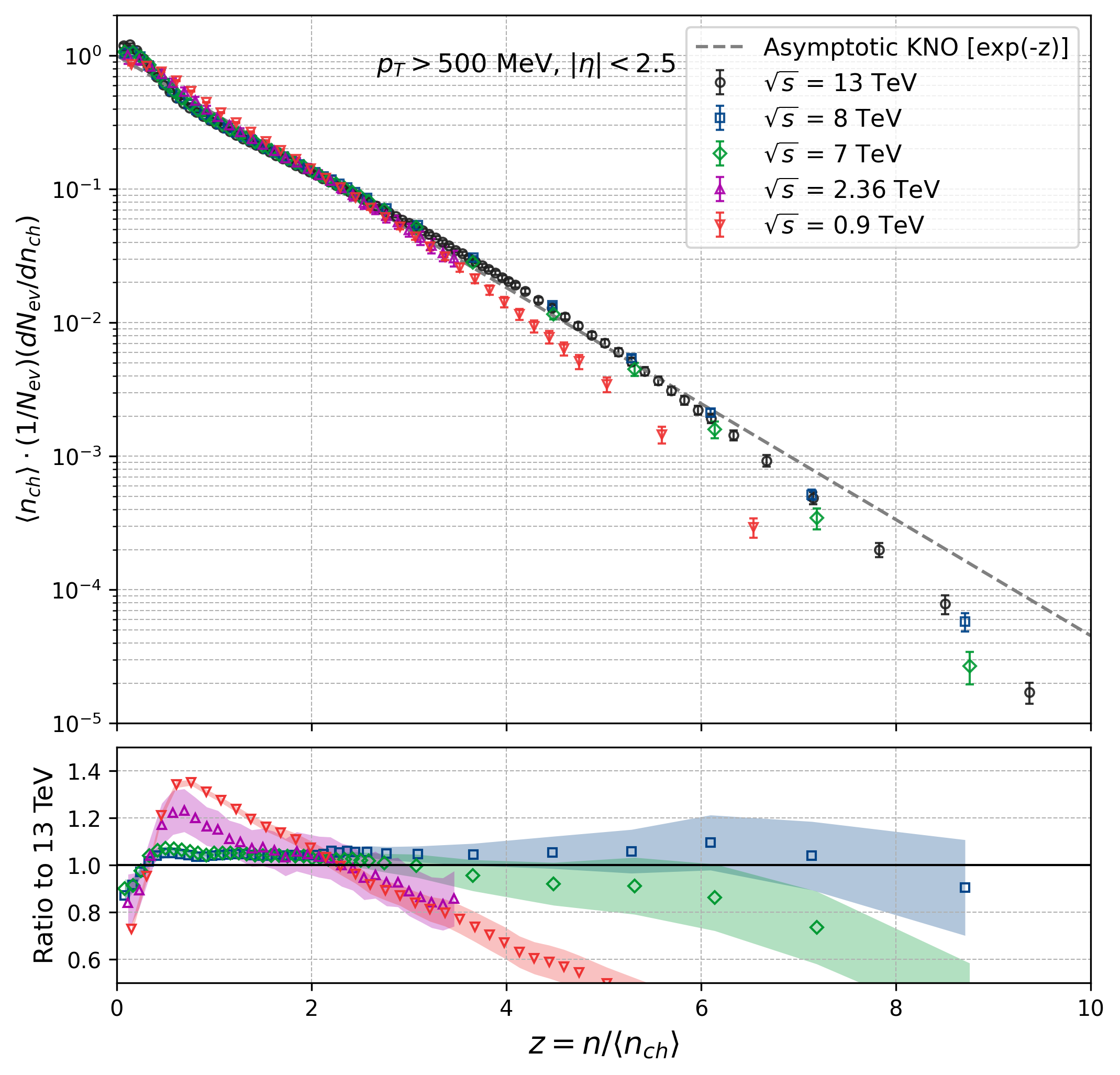}}  
(a)
\\
\end{minipage}
\hfill
\begin{minipage}[h]{0.50\textwidth} 
\center{\includegraphics[width=1.0\linewidth]{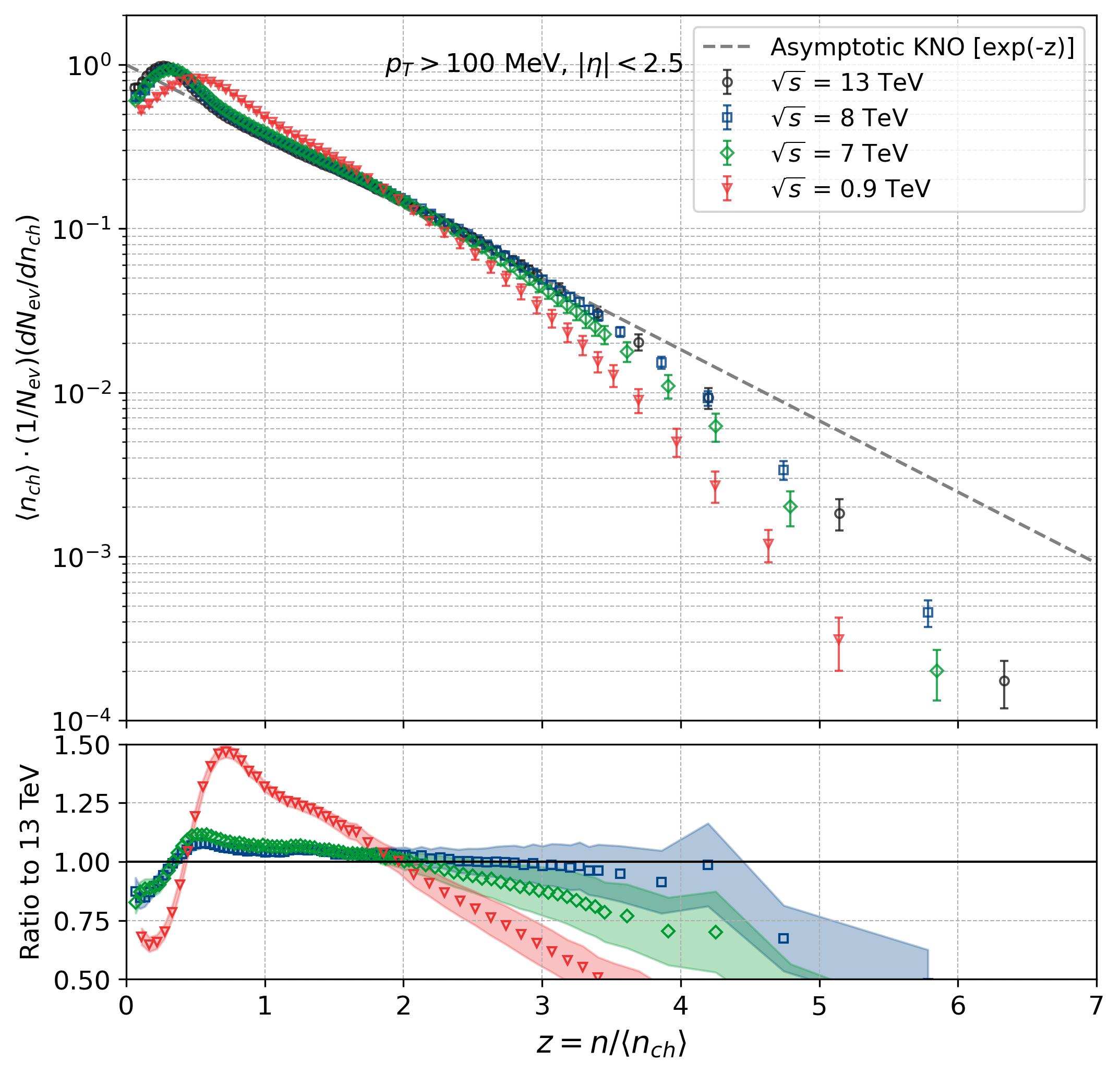}}  
(b)
\end{minipage}
\caption{ \footnotesize The graphs~\cite{Kulchitsky:2022gkm} show the  experimental data  of $\mathtt{p}\mathtt{-}\mathtt{p}$ collisions by the ATLAS 
\cite{ATLAS:2010zmc, ATLAS:2010jvh, ATLAS:2016qux, ATLAS:2016zkp, ATLAS:2016zba}  for the KNO scaled primary charged-particle  multiplicity distributions   
as a function of 
$z$ for events with  $n_{\mathrm{ch}} \ge 2$ and $\mid\eta\mid  < 2.5$ 
measurement  at the center-of-mass energies 
$0.9$, $2.36$,   $7$,  $8$ and $13$~$\mathtt{TeV}$.   The graphs show  the KNO scaled probability  $\Psi_{KNO}(z)=\langle n \rangle P_n=\langle n_{ch} \rangle (1/N_{ev}) (d N_{ev}/dn_{ch})$ as a function of the scaled  variable $z=n/ \langle n \rangle$
Each graph shows the comparison of  the asymptotic  KNO behavior $\Psi_{KNO}(z) \simeq e^{-z}$ to the experimental data at different energies for .$p_{\mathrm{T}} >500$~$\mathtt{MeV}$ (on the left) and $p_{\mathrm{T}} >100$~$\mathtt{MeV}$ (on the right) .
} 
\label{plotATLAS}
\end{figure*}

Encouraged by these findings we look for their experimental  evidence. 
The most suitable presentation of the experimental data for our purpose  is done  by Kulchitsky and  Tsiareshka~\cite{Kulchitsky:2022gkm}. They analyzed the   experimental data  of $\mathtt{p}\mathtt{-}\mathtt{p}$ collisions by the ATLAS 
\cite{ATLAS:2010zmc, ATLAS:2010jvh, ATLAS:2016qux, ATLAS:2016zkp, ATLAS:2016zba}  for the KNO scaled primary charged-particle  multiplicity distributions   
as a function of 
$z$ for events with  $n_{\mathrm{ch}} \ge 2$, $p_{\mathrm{T}} >100$~$\mathtt{MeV}$, and $p_{\mathrm{T}} >500$~$\mathtt{MeV}$ and $\mid\eta\mid  < 2.5$ 
measurement  at the center-of-mass energy 
$0.9$, $2.36$,  $7$,  $8$ and $13$~$\mathtt{TeV}$ as shown in Fig.~\ref{plotATLAS}. 

\begin{figure*}[htb!]
\begin{minipage}[h]{0.50\textwidth} 
\center{\includegraphics[width=1.0\linewidth]{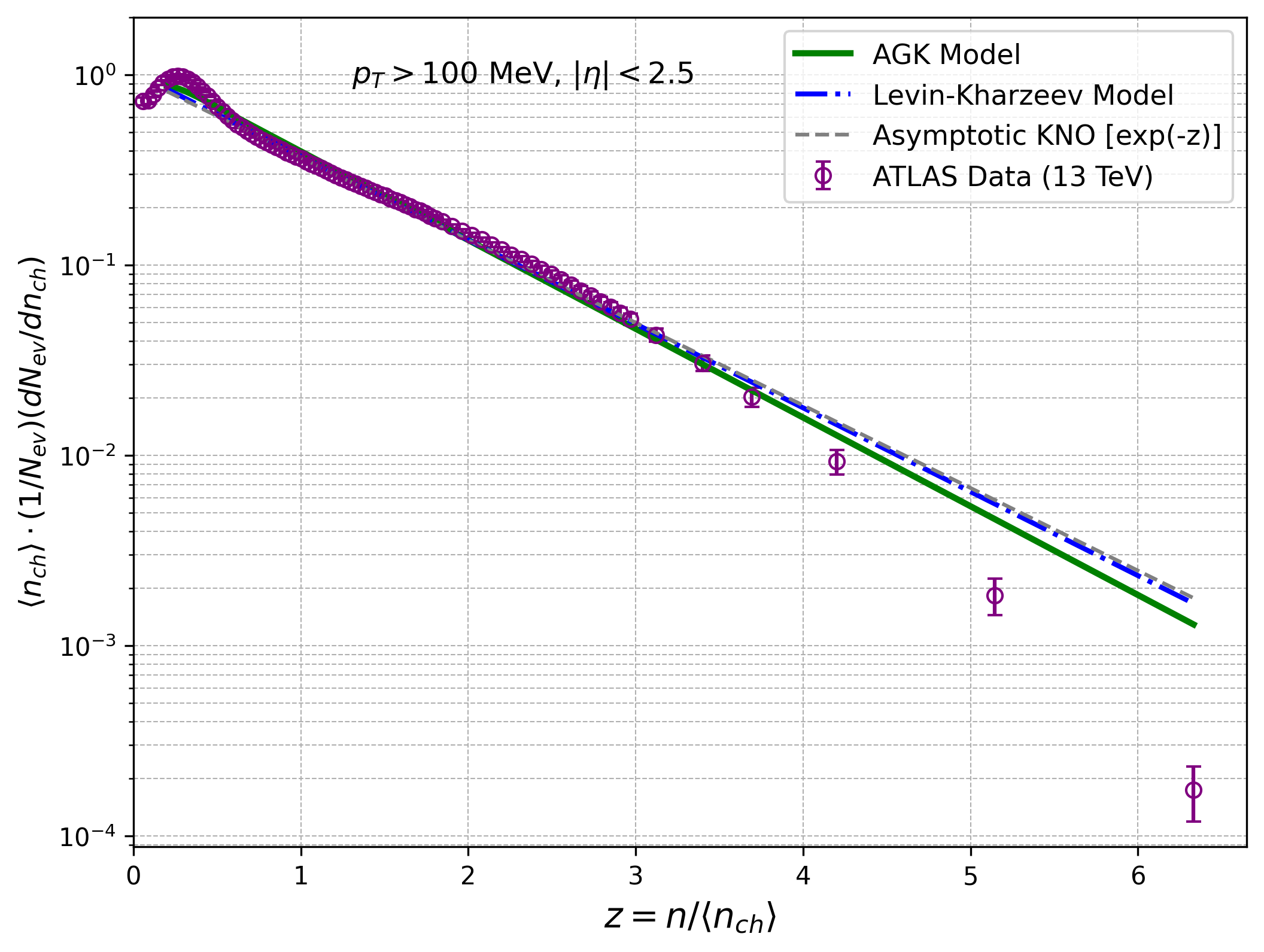}}  
(a)
\\
\end{minipage}
\hfill
\begin{minipage}[h]{0.50\textwidth} 
\center{\includegraphics[width=1.0\linewidth]{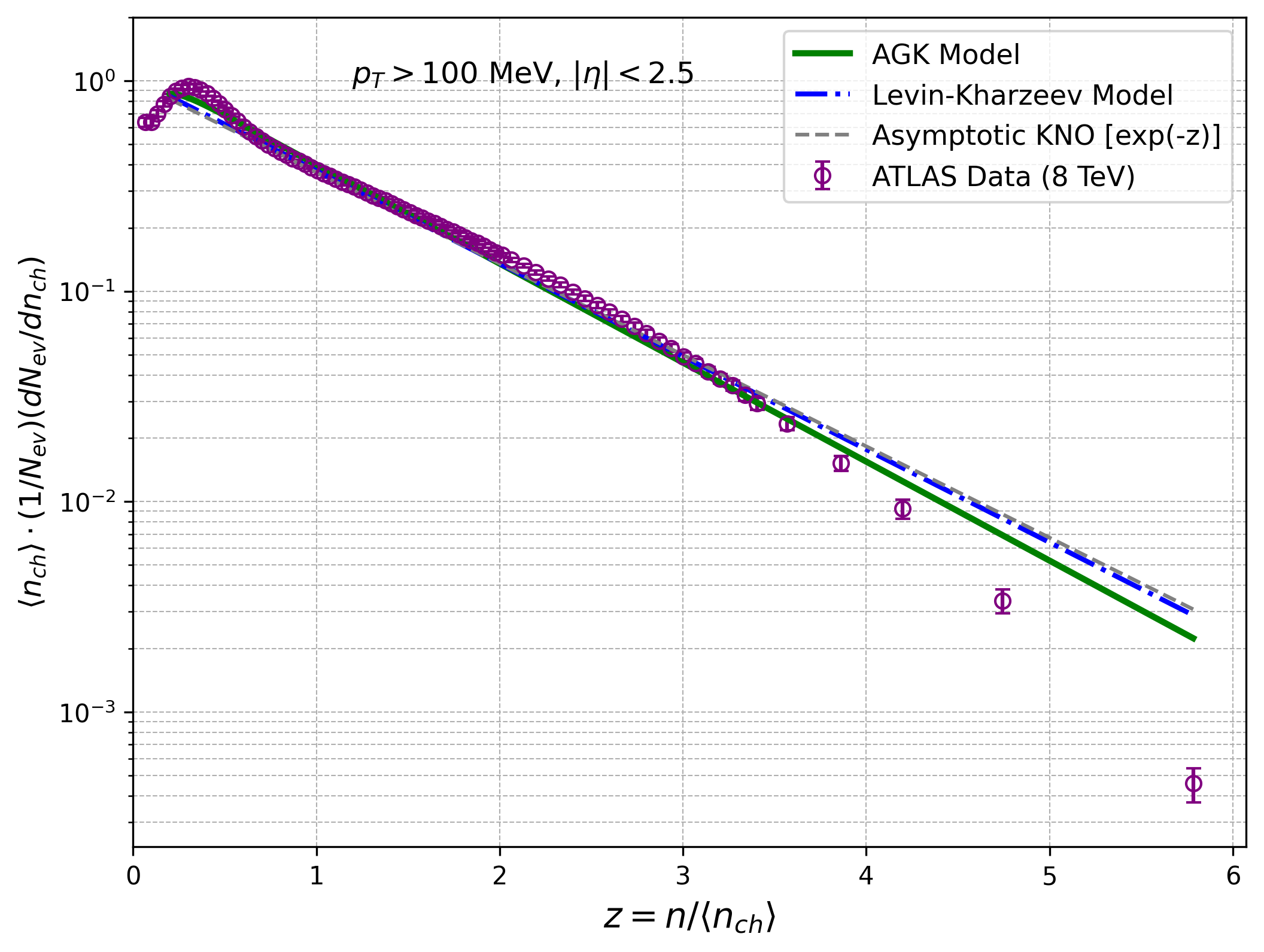}}  
(b)
\\
\end{minipage}
\begin{minipage}[h]{0.50\textwidth} 
\center{\includegraphics[width=1.0\linewidth]{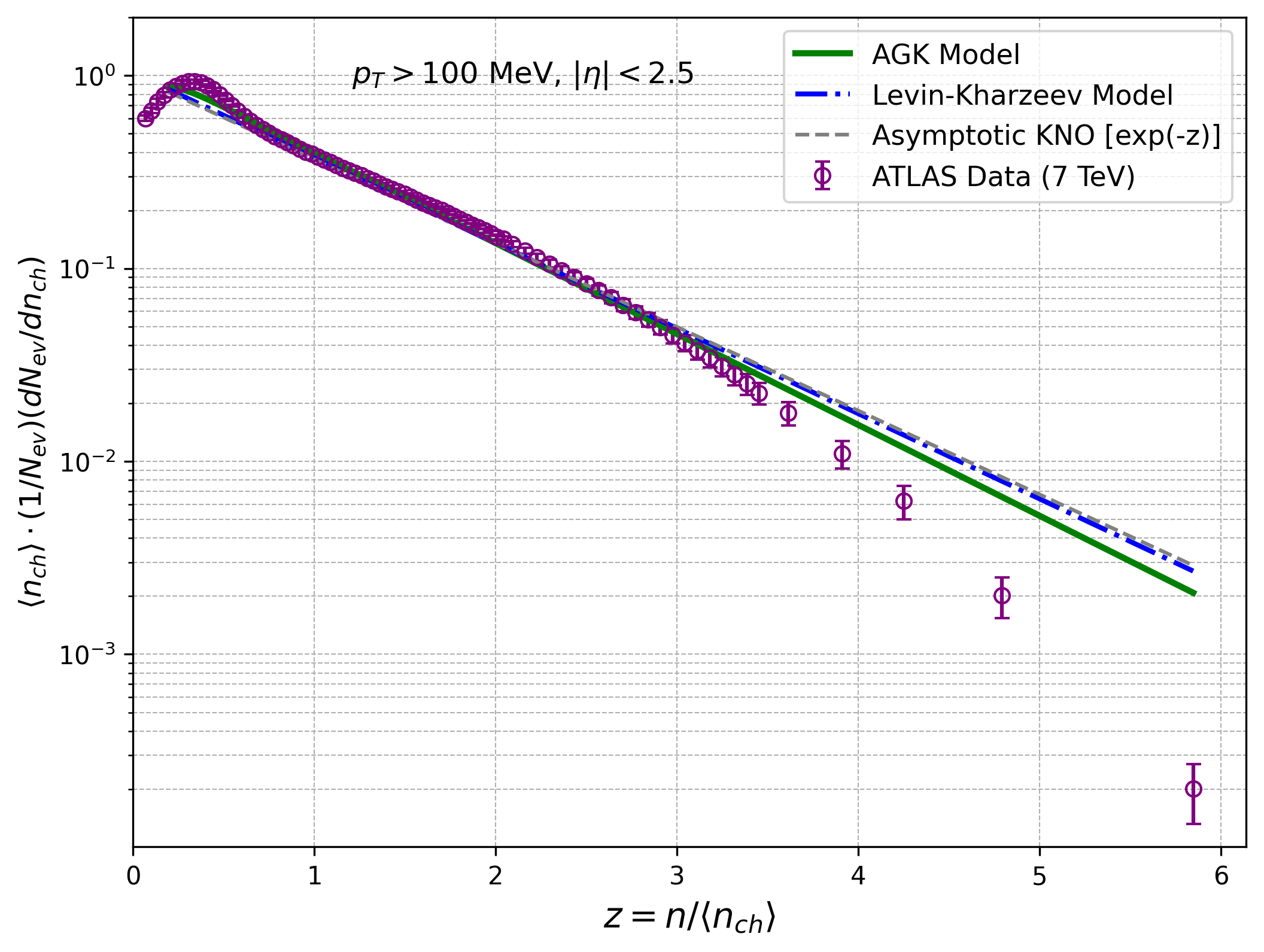}}  
(c)
\\
\end{minipage}
\hfill
\begin{minipage}[h]{0.50\textwidth} 
\center{\includegraphics[width=1.0\linewidth]{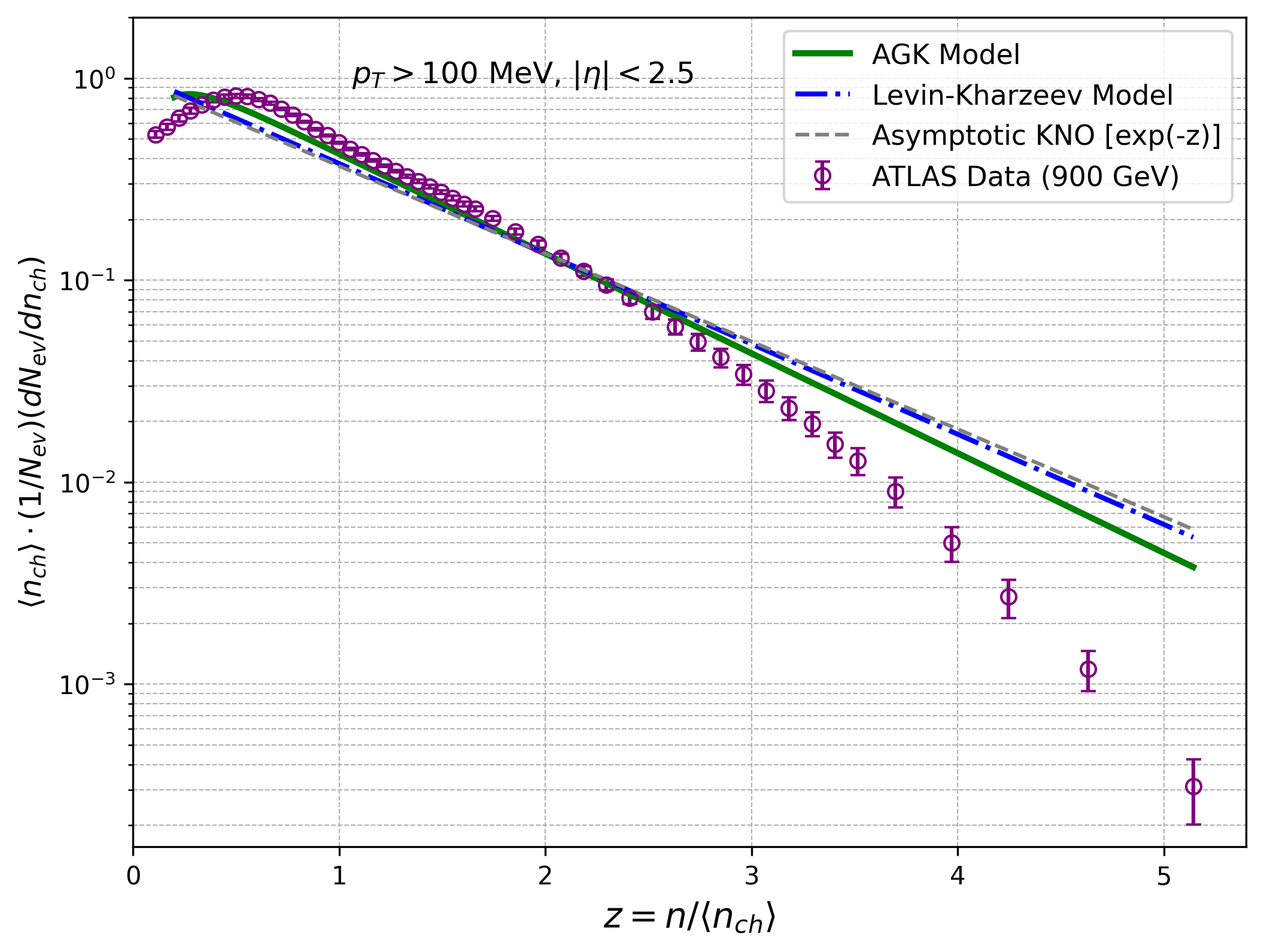}}  
(d)
\\
\end{minipage}
\caption{ \footnotesize Experimental data  of $\mathtt{p}\mathtt{-}\mathtt{p}$ collisions by the ATLAS 
\cite{ATLAS:2010zmc, ATLAS:2010jvh, ATLAS:2016qux, ATLAS:2016zkp, ATLAS:2016zba}  for the KNO scaled primary charged-particle  multiplicity distributions   
as a function of 
$z$ for events with  $n_{\mathrm{ch}} \ge 2$, $p_{\mathrm{T}} >100$~$\mathtt{MeV}$ and $\mid\eta\mid  < 2.5$ 
measurement  at the center-of-mass energies 
$0.9$,   $7$,  $8$ and $13$~$\mathtt{TeV}$.   
Each graph shows the comparison of  the Levin-Kharzeev model, the AGK model and the asymptotic  KNO behavior $e^{-z}$ to the experimental data at different energies.
All three models are very good in describing experimental data in the  $1 \leq z \leq 3$ region, while  the AGK model describes better  the experimental data for large $z$.} 
\label{plot100}
\end{figure*}

\begin{figure*}[htb!]
\begin{minipage}[h]{0.5\textwidth} 
\center{\includegraphics[width=0.9\linewidth]{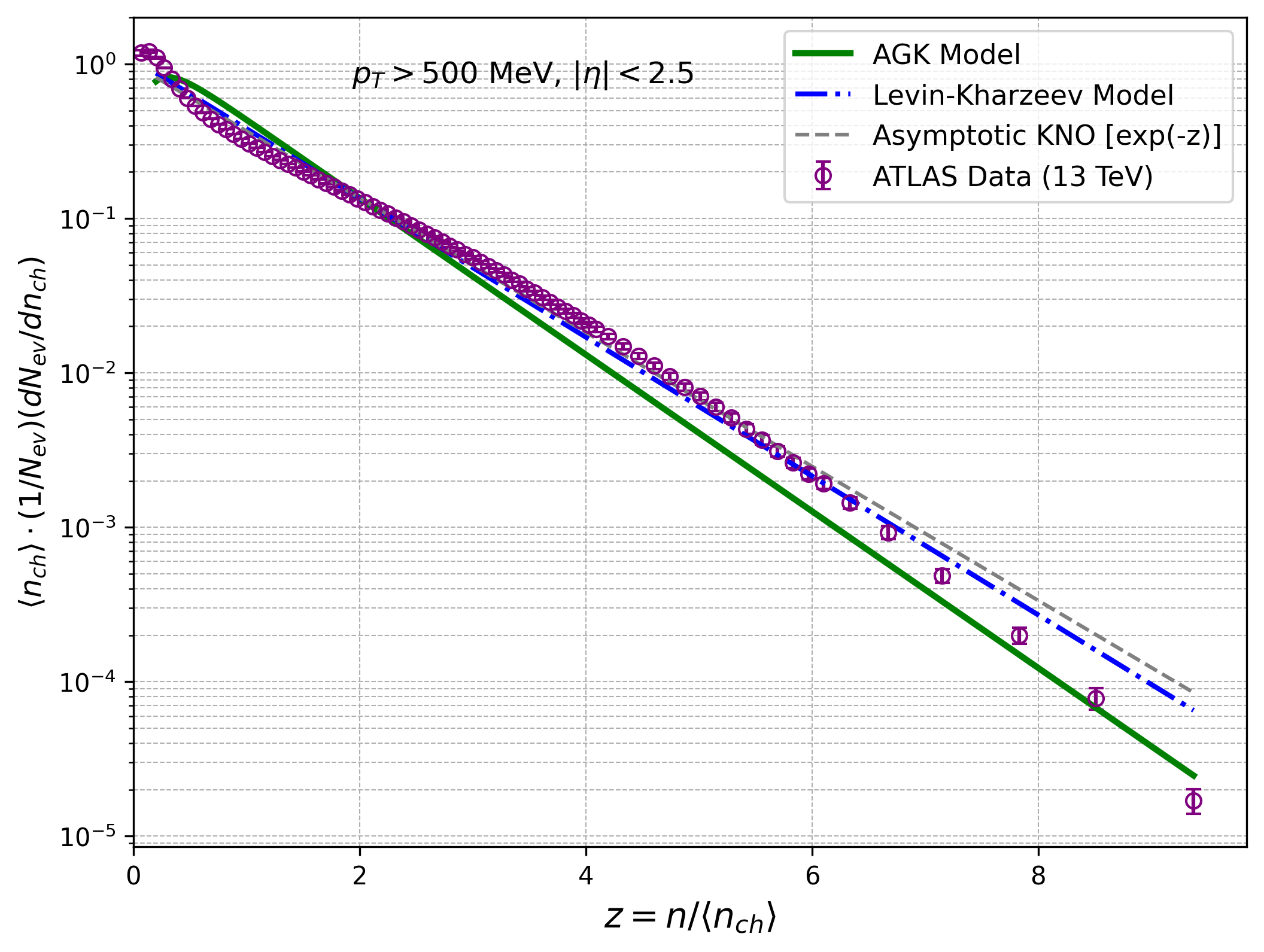}}  
(a)
\\
\end{minipage}
\hfill
\begin{minipage}[h]{0.5\textwidth} 
\center{\includegraphics[width=0.9\linewidth]{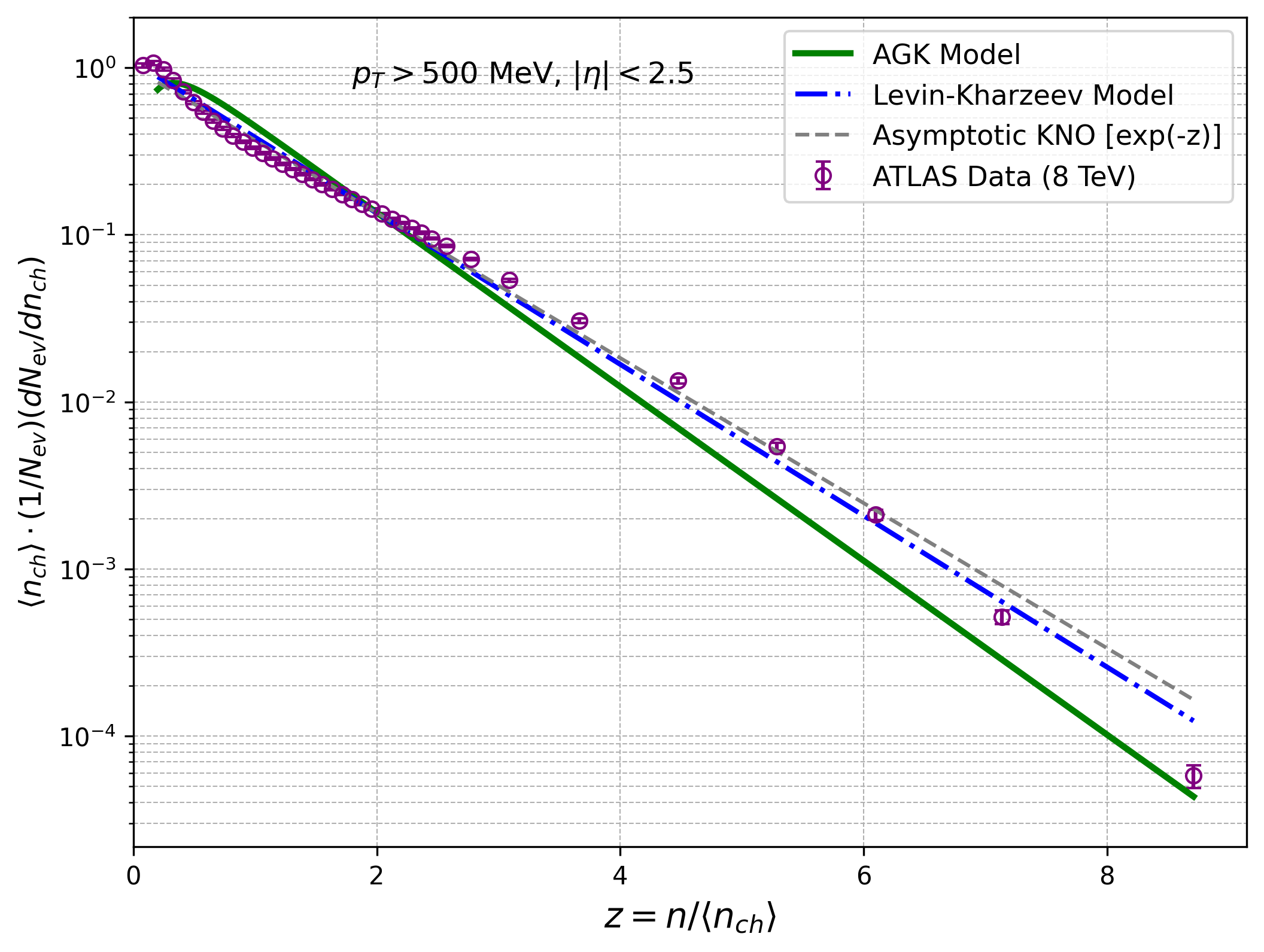}}  
(b)
\\
\end{minipage}
\begin{minipage}[h]{0.5\textwidth} 
\center{\includegraphics[width=0.9\linewidth]{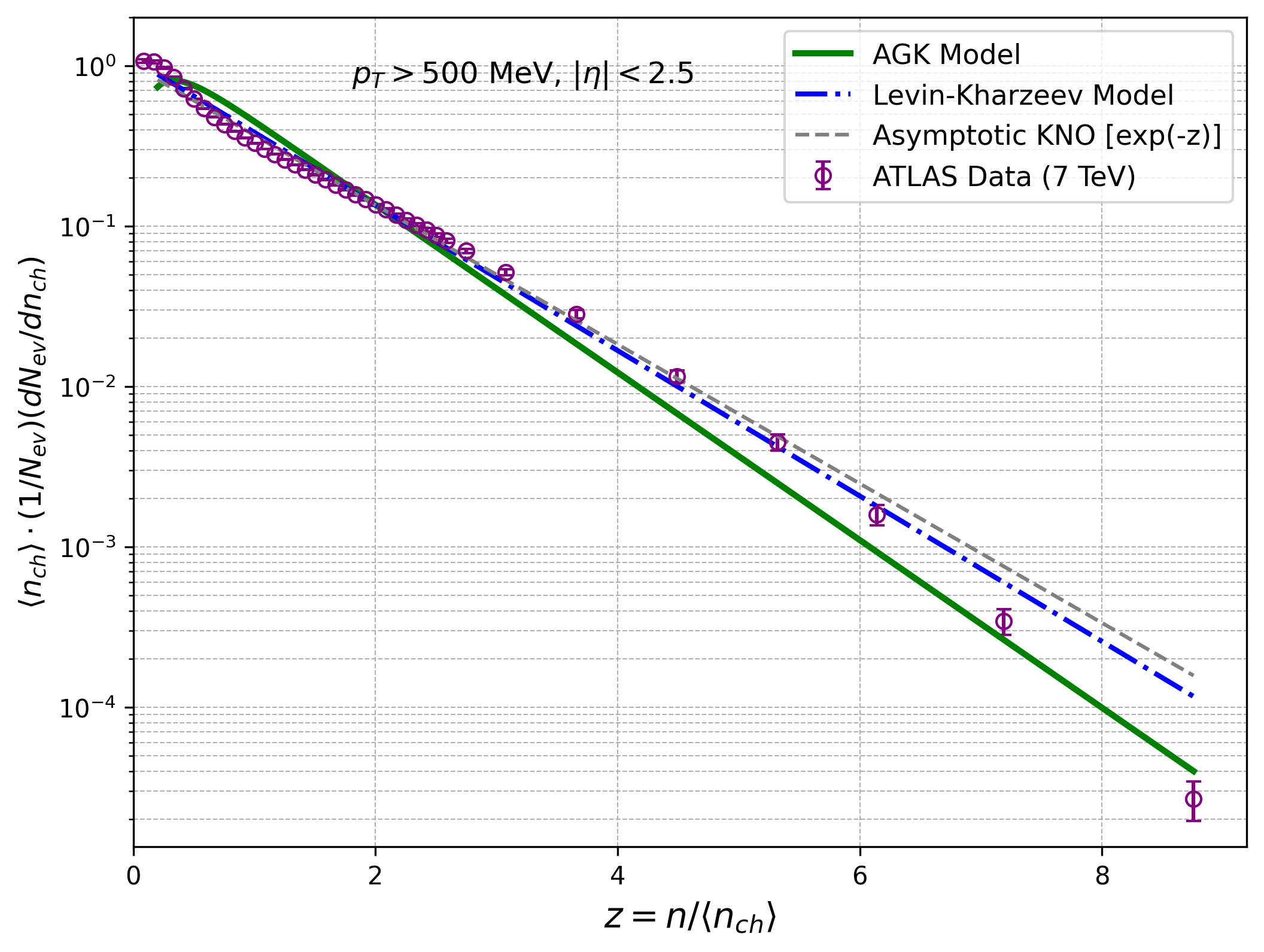}}  
(c)
\\
\end{minipage}
\hfill
\begin{minipage}[h]{0.5\textwidth} 
\center{\includegraphics[width=0.9\linewidth]{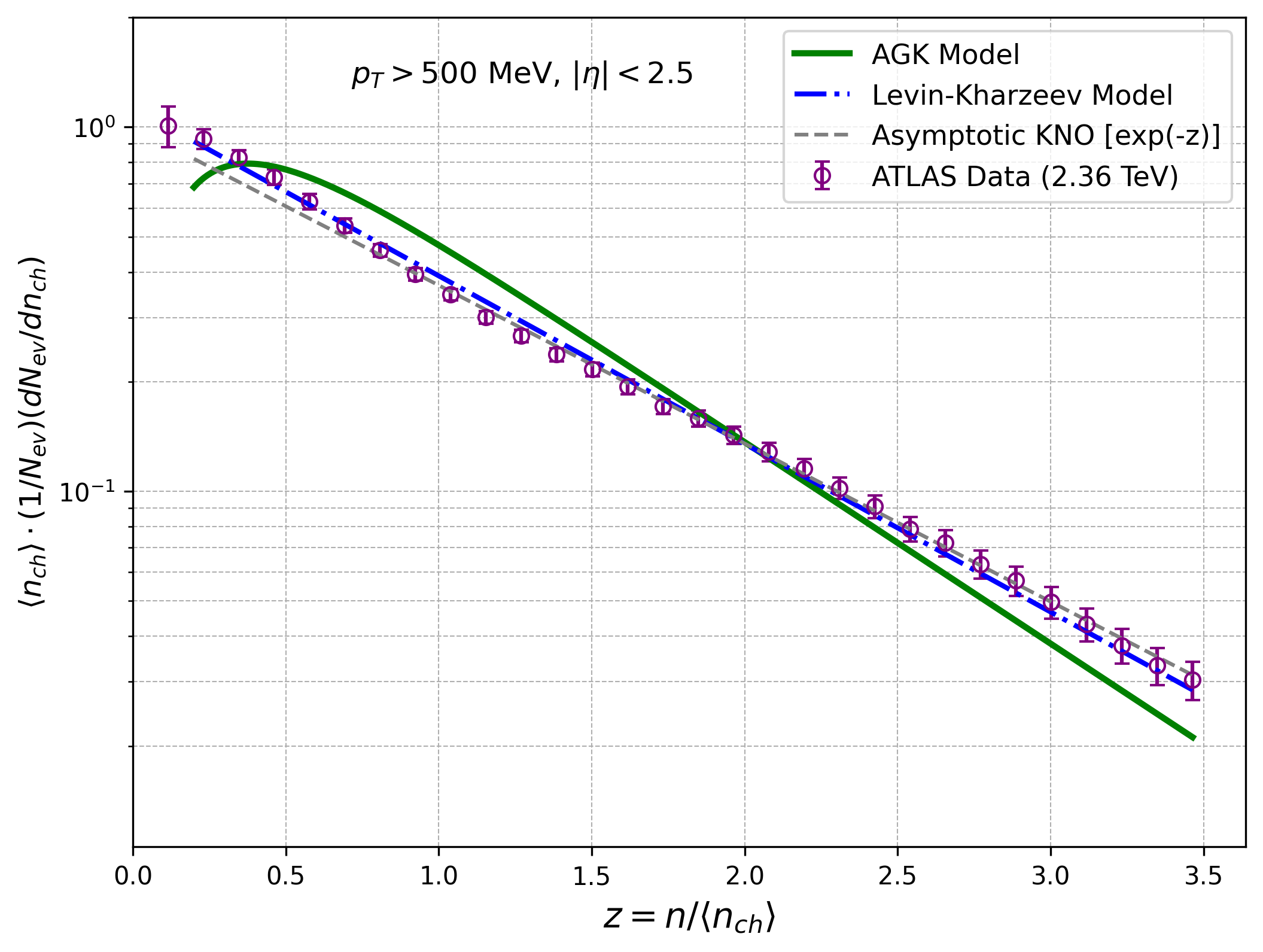}}  
(d)
\\
 \end{minipage}
 \begin{minipage}[h]{0.5\textwidth} 
\center{\includegraphics[width=0.9\linewidth]{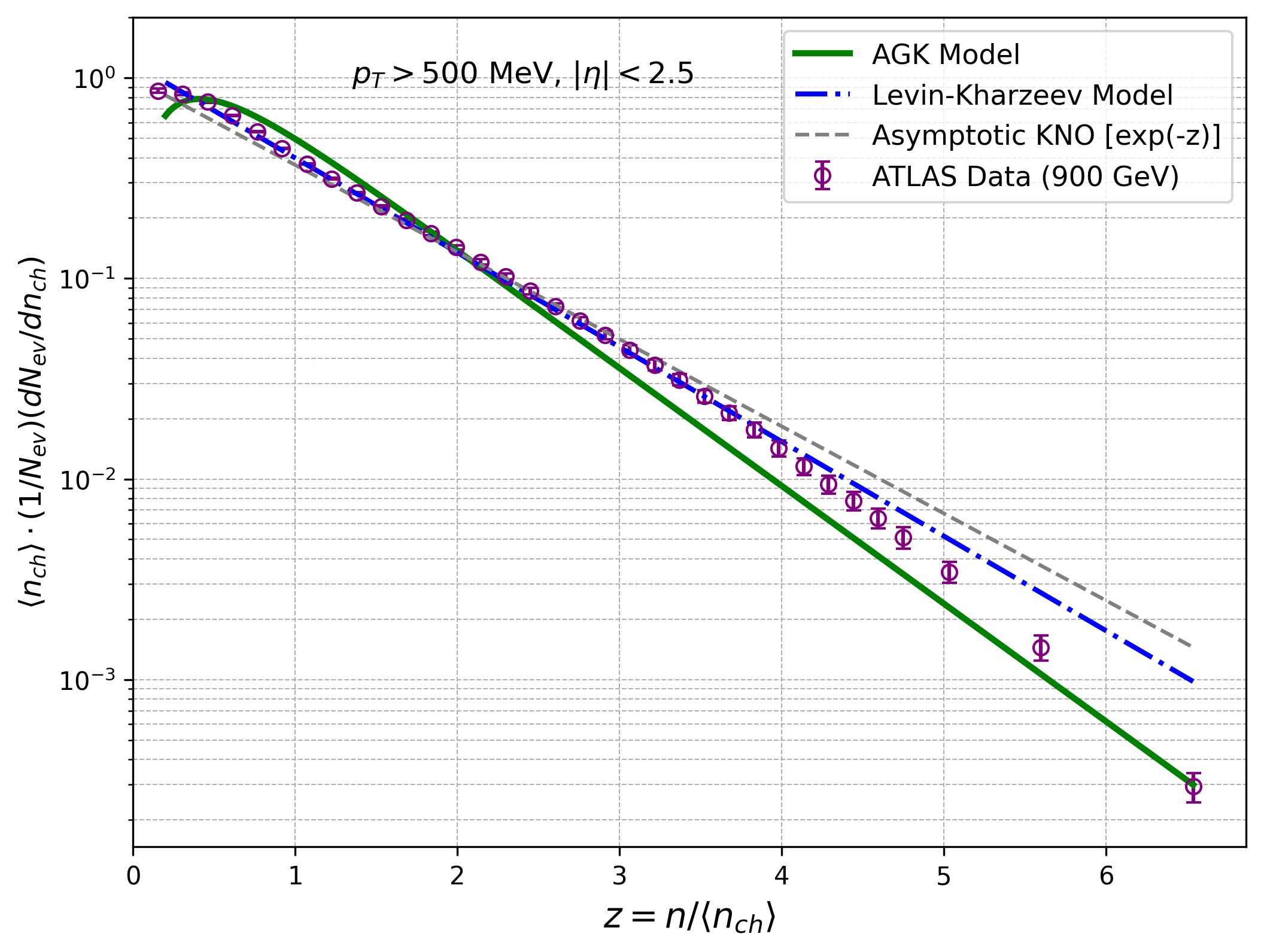}}  
(e)
\\
\end{minipage}
\hfill
\begin{minipage}[h]{0.5\textwidth} 
\center{}  
\end{minipage}
\caption{ \footnotesize Experimental data  of $\mathtt{p}\mathtt{-}\mathtt{p}$ collisions by the ATLAS 
\cite{ATLAS:2010zmc, ATLAS:2010jvh, ATLAS:2016qux, ATLAS:2016zkp, ATLAS:2016zba}  for the KNO scaled primary charged-particle  multiplicity distributions   
as a function of 
$z$ for events with  $n_{\mathrm{ch}} \ge 2$, $p_{\mathrm{T}} >500$~$\mathtt{MeV}$ and $\mid\eta\mid  < 2.5$ 
measurement  at the center-of-mass energies 
$0.9$,
   $7$,  $8$ and $13$~$\mathtt{TeV}$.   
Each graph shows the comparison of  the Levin-Kharzeev model, the AGK model and the asymptotic  KNO behavior $e^{-z}$ to the experimental data at different energies.
All three models are very good in describing experimental data in the  $1 \leq z \leq 3$ region, while  the AGK model describes better  the experimental data for large $z$.} 
\label{plot500}
\end{figure*}

The resulting graphs for   the KNO scaled data, analogous to $ \langle n\rangle P_n(\sqrt{s}) $,  show good agreement with the AGK model~\cite{Ouchen:2025tta}, the KL model and the simple asymptotic KNO scaling $e^{-z}$ for the function $\Psi(z)$. The most important outcome of our analysis is that all models describe the experimental data very well for $1 \leq z \leq 3$ and virtually coincide with the experimental values at $z=2$ in the wide range of energy    $0.9$, $2.36$,  $7$,  $8$ and $13$~$\mathtt{TeV}$ as shown in Fig.~\ref{plot500} and Fig.~\ref{plot100}. Moreover, the normalized experimental data graphs  for  $ \langle n\rangle P_n$ intersect at $z=2$   having the same value of $1/e^{2} \simeq 0.1353$ at this point as shown in Fig.~\ref{plotATLAS}. This means  the KNO violating corrections are strongly suppressed in the vicinity of $z=2$, suggesting that  it has a very special physical meaning and seems to reflect the fact that the unitarity condition (optical theorem) is saturated at $z=2$.  
At high values of $z$, the AGK model tends to outperform the KL model in the description of the experimental data. 

Indeed, the experimental data show intersection of KNO normalized probability curves  of  different energies at $z=2$ as shown in Fig.~\ref{plotATLAS}. For $p_{\text{T}}^{\min} >500 \;\mathtt{MeV}$ the  KNO normalized probability at  the   center-of-mass energies  $\sqrt{s}= 0.9 \;\mathtt{TeV}, 2.36 \;\mathtt{TeV}, 7 \;\mathtt{TeV}, 8 \;\mathtt{TeV}, 13 \;\mathtt{TeV}$ remarkably reproduce value of $1/e^2$ at $z=2$ as shown in Table~\ref{table:all500}.
 
 \begin{table}[!ht]
    \centering
    \begin{tabular}{|c|c|c|c|c|c|}
    \hline
  z     & $\mathtt{13} \mathtt{TeV}$  & $\mathtt{8} \mathtt{TeV}$  & $\mathtt{7} \mathtt{TeV}$  & $\mathtt{2.36} \mathtt{TeV}$  & $\mathtt{0.9} \mathtt{TeV}$ \\ \hline
 1.90 & 1.063 & 1.107 & 1.105 & 1.117 & 1.164 \\  \hline
 1.95 & 1.021 & 1.063 & 1.056 & 1.067 & 1.104 \\  \hline
 2.00 & 0.979 & 1.021 & 1.009 & 1.019 & 1.046 \\  \hline
 2.05 & 0.938 & 0.979 & 0.965 & 0.973 & 0.991 \\  \hline
 2.10 & 0.898 & 0.939 & 0.923 & 0.929 & 0.937 \\  \hline
\end{tabular}
       \caption{Table shows the interpolated  data of the KNO function  $\Psi_{KNO}(z) =\langle n \rangle  P_n $ for $p$-$p$ collision  of  ATLAS
Collaboration  for  the  pseudorapidity region  $-2.5 <\eta < 2.5 $, 
 the center-of-mass energies  $\sqrt{s}= 0.9 \;\mathtt{TeV}, 7 \;\mathtt{TeV}, 8 \;\mathtt{TeV}, 13 \;\mathtt{TeV}$ and $p_t > 500 \mathtt{MeV}$. The data is presented in the vicinity of $z=2$ and    normalized by $\exp(-2)$. At $z=2$, experimental values of $\Psi_{KNO}(z)$ are consistent across different energies, converging to $\exp(-2)$ within a few percent error.}
       \label{table:all500}
\end{table}

The experimental data for $p_{\text{T}}^{\min} >100 \;\mathtt{MeV}$ show a similar feature of simultaneous intersection of the KNO normalized probability curves for 
the   center-of-mass energies  $\sqrt{s}= 0.9 \;\mathtt{TeV}, 7 \;\mathtt{TeV}, 8 \;\mathtt{TeV}, 13 \;\mathtt{TeV}$. However, in contrast to the experimental data for $p_{\text{T}}^{\min} >500 \;\mathtt{MeV}$, the value  of the KNO normalized probability at the intersection point differs from $1/e^2$ by about $10 \%$. 

This model driven observation of universality of the experimental data for the KNO normalized probability at $z=2$ and its vicinity suggests the special status  of this point, where the KNO violating corrections are strongly suppressed and the KNO scaling seems to hold with accuracy of the order of experimental errors.  

Our analysis shows that the location of the point $z=2$, where the KNO violating corrections of the order of $\frac{1}{\langle n\rangle}$ vanish,  depends on the initial condition of underlying dynamics of any   \textit{memoryless} model that  respects unitarity. The KL model is exact  memoryless (geometric distribution), i.e. each iteration step is independent of the previous one.  The AGK model slightly deviates from being a  memoryless distribution, but it is  very close to the geometric distribution in the limiting case of the  large average multiplicity, i.e. high energy limit.  The initial conditions of the two models are identical, namely, at the initial rapidity the evolution starts with one countable degree of freedom, either color dipole or pomeron.

The comparison of the KL model and the AGK model with the  ATLAS experimental data for center-of-mass up to  $13 \mathtt{TeV}$ is illustrated in Fig.~\ref{plot100}  for  $p_t > 100 \mathtt{MeV}$ and in Fig.~\ref{plot500} for $p_t > 500 \mathtt{MeV}$.
 The graphs  show the KNO function  $\Psi_{KNO}(z) =\langle n \rangle  P_n $  versus the KNO variable $z=n/ \langle n \rangle$ calculated using the KL model, the 
AGK model and its asymptotic form $\exp(-z)$ compared to the ATLAS experimental data in the experimentally available range of $z$.  
The AGK model performs slightly better than the KL model especially for $p_t > 100 \mathtt{MeV}$.  All graphs for all energies and $p_t$ respective values intersect at $z=2$ having value very close to $\exp(-2)$ as shown in Table~\ref{table:KLvsAGK500all} for $p_t > 500 \mathtt{MeV}$.

\newpage

In Table~\ref{table:KLvsAGK500all} we use the original binning provided by the ATLAS collaboration without interpolating their result in the vicinity of $z=2$. In most cases the experimental data deviates by a fraction of percent from the value predicted by the AGK model in the vicinity of $z=2$.

\begin{table}[!ht]
    \centering
     \begin{tabular}{|c|c|c|c|c|c|}
    \hline
  z     & Experiment   & $e^{-z}$, $\%$  & KL model, $\%$  & AGK model, $\%$ \\ \hline
 1.90996 & 0.142774 & 3.72 & 3.97 & 5.51 \\ \hline
 1.97817 & 0.134696 & 2.69 & 2.69 & 3.27 \\ \hline
 2.04638 & 0.127601 & 1.25 & 1.0 & 0.66 \\ \hline
 2.1146 & 0.119977 & 0.59 & 0.09 & -1.14 \\ \hline
 2.18281 & 0.113117 & -0.35 & -1.08 & -3.17 \\ \hline
\end{tabular}
\\
$\sqrt{s}= 13 \;\mathtt{TeV}$
\vspace{0.5cm}
\\ 
  
 \begin{tabular}{|c|c|c|c|c|c|}
    \hline
  z     & Experiment   & $e^{-z}$, $\%$  & KL model, $\%$  & AGK model, $\%$ \\ \hline
 1.87755 & 0.152513 & 0.3 & 0.71 & 3.12 \\ \hline
 1.95918 & 0.142713 & -1.22 & -1.16 & -0.08 \\ \hline
 2.04082 & 0.133648 & -2.79 & -3.08 & -3.25 \\ \hline
 2.12245 & 0.124705 & -3.98 & -4.6 & -5.98 \\ \hline
\end{tabular}
\\
$\sqrt{s}= 8 \;\mathtt{TeV}$ 
\vspace{0.5cm}
\\
\begin{tabular}{|c|c|c|c|c|c|}
    \hline
  z     & Experiment   & $e^{-z}$, $\%$  & KL model, $\%$  & AGK model, $\%$ \\ \hline
 1.91987 & 0.146995 & -0.25 & -0.02 & 1.75 \\ \hline
 2.00334 & 0.135733 & -0.63 & -0.77 & -0.33 \\ \hline
 2.08681 & 0.126748 & -2.1 & -2.6 & -3.47 \\ \hline
 2.17028 & 0.117212 & -2.62 & -3.47 & -5.59 \\ \hline
\end{tabular}
\\
$\sqrt{s}= 7 \;\mathtt{TeV}$ \vspace{0.5cm}
\\
\begin{tabular}{|c|c|c|c|c|c|}
    \hline
  z     & Experiment   & $e^{-z}$, $\%$  & KL model, $\%$  & AGK model, $\%$ \\ \hline
 1.84758 & 0.158565 & -0.6 & 0.11 & 4.49 \\ \hline
 1.96305 & 0.142717 & -1.6 & -1.62 & 0.25 \\ \hline
 2.07852 & 0.128255 & -2.45 & -3.17 & -3.68 \\ \hline
 2.194 & 0.115351 & -3.36 & -4.77 & -7.55 \\ \hline
\end{tabular}
\\
$\sqrt{s}= 2.36 \;\mathtt{TeV}$
\vspace{0.5cm}
\\
\begin{tabular}{|c|c|c|c|c|c|}
    \hline
  z     & Experiment   & $e^{-z}$, $\%$  & KL model, $\%$  & AGK model, $\%$ \\ \hline
 1.83767 & 0.16769 & -5.07 & -4.19 & 1.81 \\ \hline
 1.99081 & 0.143007 & -4.49 & -4.85 & -2.7 \\ \hline
 2.14395 & 0.12087 & -3.04 & -4.67 & -6.25 \\ \hline
\end{tabular}
\\
$\sqrt{s}= 0.9 \;  \mathtt{TeV}$
\\
 \caption{The table shows the deviations of the KL mode, the AGK model and the exponential KNO function from the ATLAS experimental data for $p-p$ collisions (in percents) for $|\eta|< 2.5$, $p_t > 500 \;\mathtt{Mev}$ and center-of-mass energy between $0.9\; \mathtt{TeV}$ and $13\; \mathtt{TeV}$. }
        \label{table:KLvsAGK500all}
\end{table}

The  point $z=2$ implies that the number of produced particles $n$ equals twice the average multiplicity $\langle n \rangle$, i.e. $n =2 \langle n \rangle$. At $z=2$ the KNO scaling seems to hold for a wide range of  energies as the KNO violating corrections are suppressed by the order of $1/\langle n \rangle^2$, which is comparable to the experimental error.  At  $z=2$ all experimental curves for the KNO scaled probability $ \langle n \rangle P_n$ intersect and have value $1/e^2$ that originates from the memoryless geometric distribution of the produced particles that has the maximal entanglement  entropy. 
This suggests that the  memoryless geometric distribution  saturates $n =2 \langle n \rangle$, which is closely related to the optical theorem $\sigma= 2 \Im A$.  The combination of these two observations establishes a direct relation between the optical theorem and the maximal entanglement  entropy.

\section{Conclusion and Discussions}\label{}
 In this paper we  expand the KNO scaled probability $ \langle n \rangle P_n$ at large values of the average multiplicity of the produced particles $\langle n \rangle$. The expansion of $ \langle n \rangle P_n$ for the probability  $P_n$ calculated using the  Markov chain with AGK cutting rules  reveals a special point of the KNO scaled variable $z=n / \langle n \rangle=2$,  where the first order corrections in the inverse powers of $\langle n \rangle$ vanish. The higher order terms in the expansion of the KNO scaled probability $ \langle n \rangle P_n$  are  of the order of the experimental error for $\mathtt{p-p}$ collision  ATLAS experiment  at $1\; \mathtt{TeV}$ and above.  We show that without any reference to the theoretical model used, the experimental curves for $ \langle n \rangle P_n$ cross at $z=2$ in a wide range of experimentally available  energies up to $13\; \mathtt{TeV}$.
 The universality found in the analytic calculations and supported by the proper analysis of the experimental results suggests a special status of $z=2$. 
 The detailed analysis of the theoretical model shows that this universality is related to the fact that in the high energy limit $P_n$ approximates geometrical distribution, which is the only memoryless discrete distribution. We claim that every model that satisfies the unitarity and the same initial condition we used, must reproduce the universal scaling point $z=2$ in the high energy limit, i.e. the limit of the large average multiplicity of the produced particles $\langle n \rangle$. We argue that $z=n / \langle n \rangle=2$ implies statistical saturation of the optical theorem $\sigma =2 \Im A$ by the stochastic process, where $\sigma \simeq n$ and $\Im A\simeq \langle n \rangle$.  Our analysis adds to the recent advances~\cite{Baker:2017wtt,Kharzeev:2021nzh,Hentschinski:2022rsa,Kutak:2023cwg,Hentschinski:2023nhy,Hentschinski:2023izh,Hentschinski:2024gaa,Grieninger:2023ufa,Chachamis:2023omp,Bhattacharya:2024sno,Ouchen:2025tta,Grieninger:2025wxg,Datta:2024hpn,Hentschinski:2025pyq,Kutak:2025hzo,Kutak:2025syp,Fucilla:2025kit,Moriggi:2025qfs,Hatta:2024lbw,Ramos:2025tge,Moriggi:2024tiz} in understanding the hadronization in the high energy scattering physics using quantum entanglement and von Neumann entropy methods adopted from  the quantum information science.

\section{Acknowledgement}\label{}

We are indebted to Sergey Bondarenko  for inspiring discussions on the topic. 
This work is supported in part by "Program of HEP support- Council of Higher Education of Israel".


\end{document}